\def\year{2015}
\documentclass[letterpaper]{article}
\usepackage{aaai}
\usepackage{times}
\usepackage{helvet}
\usepackage{courier}
\usepackage{hyperref}
\usepackage{graphicx}
\begin{document}
%
\title{A Python Engine for Teaching Artificial Intelligence in Games}
\author{Mark O. Riedl\\
School of Interactive Computing\\
Georgia Institute of Technology\\
riedl@cc.gatech.edu
}
\maketitle
\begin{abstract}
\begin{quote}
Computer games play an important role in our society and motivate people to learn computer science. 
Since artificial intelligence is integral to most games, they can also be used to teach artificial intelligence.
We introduce the Game AI Game Engine (GAIGE), a Python game engine specifically designed to teach about how AI is used in computer games.
A progression of seven assignments builds toward a complete, working Multi-User Battle Arena (MOBA) game.
We describe the engine, the assignments, and our experiences using it in a class on Game Artificial Intelligence.
\end{quote}
\end{abstract}


\section{Introduction}

\noindent 

For many, exposure to computer games leads to an interest in computer programming and computer science.
Computer games are also a gateway for interest in artificial intelligence; computer games are often the first exposure people have to AI.
Artificial intelligence in computer games appears simple, when compared to the state of the art in AI research.
However, this also makes computer games an excellent way to introduce people to artificial intelligence programming.
The virtues of using computer game environments to teach computer programming have been well-explored \cite{denero10,wong10,taylor11,sosnowski13}. 
Computer games have also been used to teach artificial intelligence, including the Berkeley Pac-Man AI course materials (\url{http://ai.berkeley.edu/project_overview.html}).

Computer game artificial intelligence, often shorted as {\em Game AI}, is often considered a distinct sub-discipline of artificial intelligence focused on the short-term illusion of believable agent behavior. 
In computer games, artificial intelligence in some form is responsible for the behavior for every virtual entity encountered by the human player.
In some games, some form of artificial intelligence may also take additional roles such as generating terrain or adapting the game to increase player engagement. 
Many of the algorithms taught in introductory artificial intelligence are prominently used in creating engaging virtual experiences: graph search, agent decision-making, planning, and, to a lesser extent, machine learning and data analysis. 

Game AI, however, focuses on pragmatic aspects of artificial intelligence as applied to the highly-constrained computing environment of computer games.
Game AI must be concerned with scalability. Virtual agents must operate in real-time and there can be dozens, hundreds, or thousands of agents consuming a highly-constrained number of computation cycles.
Game AI is also concerned with player experience. It explores the question of what behaviors can be conducted by a virtual agent facilitate players' momentarily suspension of disbelief.
Often the constraints of scalability and player experience result in ``simple'' solutions such as finite state machines, which belie the complexity of design need to create the illusion of behavior that would otherwise be considered ``AI Hard'' in real-time with few computational resources.
Consequently, many universities teach Game AI as a separate course from an introductory artificial intelligence course.
Game AI classes emphasize design and selecting the right AI technique for the job.

Because AI is an integral part of any game design, the AI algorithms cannot be easily separated from the game.
In this paper, we introduce a game engine specifically designed for teaching Game AI.
Instead of asking students to integrate AI into an existing game, the {\em Game AI Game Engine} (GAIGE) provides all the core functionality of a computer game except for the AI algorithms.
Through a controlled progression of Python programming assignments, students build a fully functional Multi-User Online Battle Arena (MOBA) game piece by piece.
Each piece requires the implementation of a Game AI algorithm.
The progression breaks the task of creating a fully functioning computer game into manageable pieces that can be individually graded by {\em autograder}, and provides students with a sense of accomplishment.
GAIGE can be downloaded from \url{http://game-ai.gatech.edu}.
 

\section{Game AI}

The term, {\em Game AI}, has come to refer to the set of tools---algorithms and representations---developed specifically to aid the creation and management of interactive, real-time, digital entertainment experiences. 
While games are played by humans, there are a number of aspects of the game playing experience that must be automated: roles that would be best performed by humans but are not practical to do so:

\begin{itemize}
\item Opponents and enemies that are meant to survive for only a short time before losing.
\item Non-player characters in roles that are not ``fun'' to play such as shopkeepers, farmers, victims, or footsoldiers.
\item Companions in single-player experiences and non-player characters in support roles.
\item Drama manager to adjust the game plot in response to real-time player behaviors \cite{riedl:aimag2013}.
\item Game designer for personalized experiences at scale.
\end{itemize}

\noindent
As we go down this list, Game AI is charged with taking progressively more responsibility for the quality of the human players' experiences in the game.

Game AI programming is often considered part of the game design process because it both constrains the creative process of designers and also realizes the creative vision of the game designers.
Because of this, {\em Game AI} is often considered a separate discipline from that that we conventionally refer to as {\em artificial intelligence}.
In artificial intelligence, rational behavior and optimality are idealized criteria for success; researchers seek to improve rationality and accuracy via increasingly sophisticated algorithms and data sets.  
In Game AI, the goal is to create the most engaging experience possible for human players.
Any and all techniques or algorithms that result in the temporary suspension of disbelief when interacting with virtual entities are valid solutions.
Furthermore, graphical rendering often consumes a vast majority of the computing power of the computing device (PCs, consoles, mobile devices, etc.), requiring any AI technique implemented within a game to be execute in near real-time and with an extreme dearth of computing cycles.
Despite the difference in goals between conventionally-defined artificial intelligence and Game AI, they draw from the same basic desire to create artificial agents that appear, at least for a short time, to perform behaviors that one might deem to require intelligence when performed by a human.

A vast majority of AI problems in game development fall into two categories: pathfinding and decision-making. 
Pathfinding is the problem of identifying an efficient path from one part of a virtual environment to another whilst avoiding collisions with obstacles that can shatter the illusion of intelligence.
Graph search algorithms such as Djikstra's Algorithm, the Floyd-Warshall algorithm, and A*, are commonly implemented in games.
Decision-making involves choosing and executing agent behaviors that enhance the engagement of the player.
Even goals that are adversarial in nature, e.g., attacking the player's avatar, may not be as straightforward as optimizing the amount of damage inflicted to the player's avatar. 
For example, different types of adversaries are expected to manifest different ``personalities'' (e.g., a zombie versus a soldier) and engage the player in different levels of challenge.

One of the most common techniques for implementing decision-making is the finite-state machine, which provides a high degree of designer control over the real-time behavior of a virtual entity and consumes little computational overhead.
Recall that they goal of Game AI is the {\em illusion} of intelligence for a short period of time, and a well-designed finite-state machine can achieve this goal effectively.
For more dynamic entity behavior, some games implement {\em behavior trees}, a greedy form of {\em hierarchical task network} (HTN) planners (c.f., \cite{ghallab04}) that create plans in real-time but cannot backtrack, bind variables, nor interleave methods.
Like finite-state machines, behavior trees provide designers with the ability to specify proper behavior while leaving the timing and, to some degree, the sequence of decisions to the agent.

Game AI includes forms of intelligence other than agent decision-making and navigation.
{\em Procedural content generation} is the use of algorithms to create game levels, maps, enemies, quests and missions. 
{\em Player modeling} is used to predict player behavior or preferences, which can then be used to alter parameters of game play---such as difficulty---or to generate content.
Data mining is used extensively to analyze how players are interacting with the game; data can be used to guide the development of patches, new content, or make business decisions.


\section{Related Work}

A course in Game AI may be taught as an advanced artificial intelligence class or as an alternative to an introductory artificial intelligence class.
One of the primary challenges of designing a Game AI course is the creation of meaningful project work that allows students to get hands-on experience making design decisions and writing AI algorithms.
There are three common approaches to coursework: 
(1) using one or more commercial computer games; 
(2) using a commercial game engine; 
or
(3) building a home-grown game and/or game engine.

Many commercially available computer games can be ``modded,'' allowing custom code to be incorporated into the game. 
However, these games prioritize graphics rendering and are not designed with any functionality in mind other than that delivered to the user.
This makes it difficult to craft meaningful pedagogical experiences because they may not expose the right functionality through APIs.
Games are likely to already have implementations of algorithms one wishes the student to write, which must be removed or ignored.
One of the biggest challenges is the substantial time that must be devoted to learning the complex underlying codebase.
In the event that any one game doesn't support all of the AI techniques covered in a class, switching between different code bases and programming languages can result in significant overhead that does not contribute directly to the pedagogical goals of the class.
Furthermore, students may be required to purchase the game, and end user licenses may impose on the rights of students to own their own coursework.

An alternative is working with general-purpose game engines, which are not full games, but provide core functionalities found across many games.
One of the most popular, free game engines is {\em Unity3D} (\url{http://unity3d.com}).
Game engines operate as high-level programming environments with specialized libraries for common aspects of games.
Commercial game engines have high learning curves because of their general purpose nature---most of the available complexity and functionality is irrelevant to the task of learning to program AI algorithms.
Further, game engines are not themselves games require a game to be created either by the instructor or by students.


Homegrown computer game environments and simulations have been developed by instructors.
The Berkeley Pac-Man AI course materials have been adopted by many AI courses.
However, Pac-Man is a toy problem in the sense that agent control of the Pac-Man agent is an artificial task---Pac-Man is supposed to be controlled by the human. 
The SEPIA simulation environment \cite{sosnowski13} is a more realistic example because the opponent in a Real Time Strategy (RTS) game is often computer-controlled.
Homegrown game and simulation environments overcome many of the limitations of commercial games and game engines.
However, students are still often asked to re-implement existing AI functionality, which may be unsatisfying and perceived as make-work.


\section{The Game AI Game Engine (GAIGE)}


Our approach follows the homegrown game engine strategy.
The {\em Game AI Game Engine} (GAIGE) is implemented in Python using the PyGame package for sprite rendering and animation. 
As a game engine, GAIGE provides general functionality used across many types of games {\em except} artificial intelligence.
The game engine, however, is designed to support Game AI instruction by providing convenient hooks for artificial intelligence implementations.

GAIGE uses simple sprite-based graphics to deemphasize the focus on the graphics of the game and keep the codebase simple enough that one can learn how it works quickly. 
For the purposes of most AI implementation in computer games, whether graphics are 3D or 2D has little bearing. 
Despite the simplicity of the code base, the game engine is modeled after much more complex engines such as the {\em Unreal Tournament} engine.
The code base is highly object-oriented and modular, which allows new functionality to be integrated into the engine by sub-classing basic components. 
For example, new game types can be created by extending a single class type.
Similarly, different pathfinding algorithms can be implemented by extending existing class types.

GAIGE is implemented in Python, a popular scripting language that is becoming a popular language for University courses.
Because GAIGE is implemented as a scripting language that is interpreted at run-time, it allows us to easy develop autograders for programming assignments.
Unlike game development courses that involve a lot of design, Game AI is an algorithms class and algorithms either produce the right result or not and operate optimally or not.
Special autograder scripts have been written that unit-test student-written code without all of the baggage of the rest of the game engine.
When the entire game engine is necessary for evaluation, the game engine can be operated in ``headless'' mode, meaning graphics are suppressed and the engine can be run many times faster than real-time.

We designed and implemented a series of seven programming assignments. 
Artificial intelligence algorithms appear prominently and numerously in computer games.
Each programming assignment contributes to the development of a fully functional {\em Multi-user Online Battle Arena} (MOBA) game.
Assignments build off each other and create a sense of progression toward an ultimate purpose, instead of a series of disjoint exercises. 
 
\begin{figure}
\includegraphics[width=3.5in]{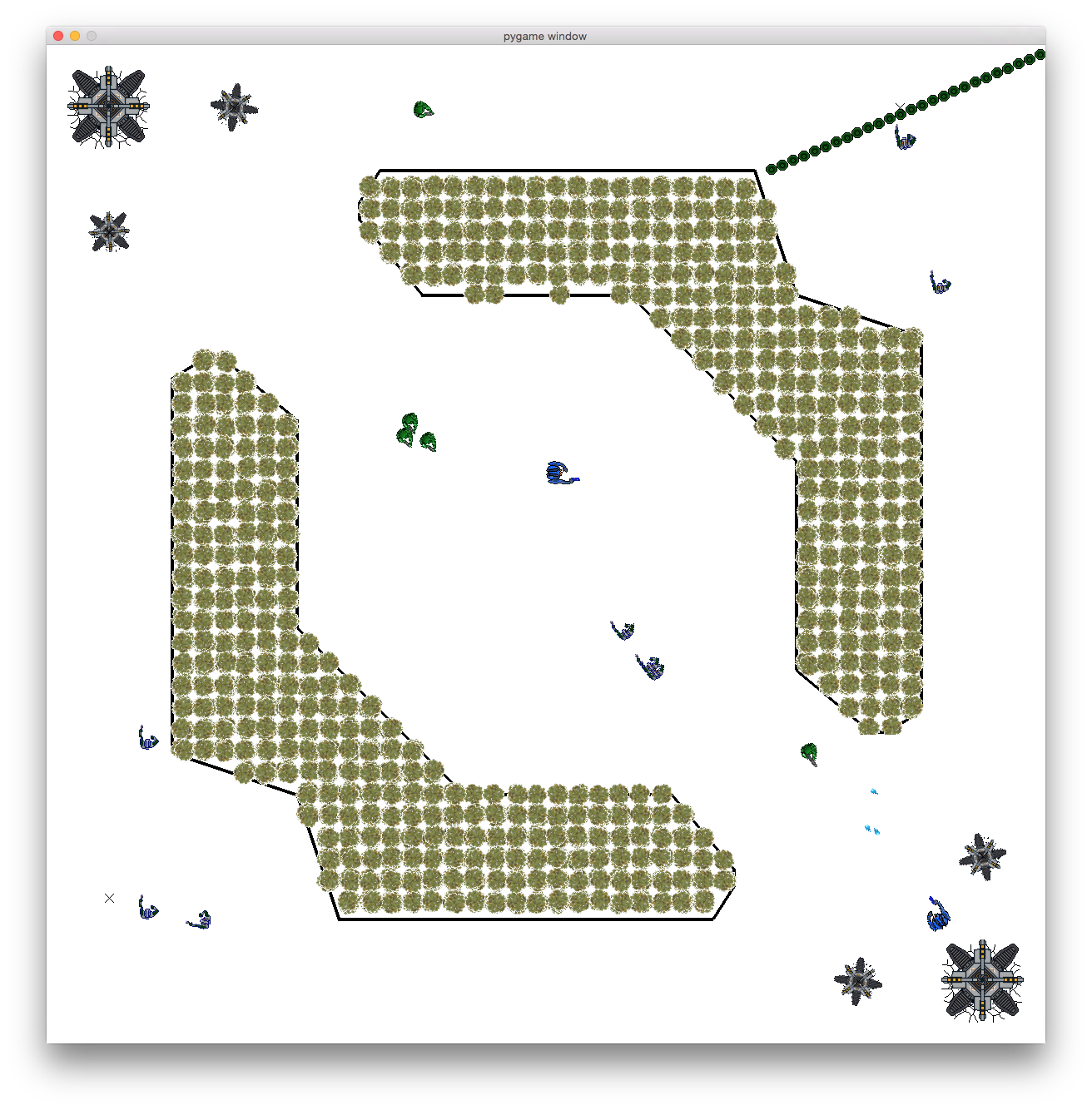}
\caption{Screenshot of GAIGE running a complete implementation of a Multi-User Online Battle Arena game.}
\label{fig:gaige}
\end{figure}


\subsection{Multi-User Online Battle Arenas}

A Multi-user Online Battle Area (MOBA) is a genre of computer game in which players on opposing teams attempt to destroy each others' bases. 
MOBAs are typically top-down perspective games and aesthetically resemble real-time strategy (RTS) games. 
Unlike RTS games in which each player is in charge of micro-managing an entire army of entities, MOBA players control a single entity, called a {\em hero}, which is just one of many entities in an army.
Entities that are not directly controlled by the player, called {\em minions}, are controlled by the computer. 
Minions are numerous, simple, and weak, whereas there are few heroes that are relatively powerful and have special abilities.
Thus, MOBAs combine elements of real-time strategy games and first-person shooters. 

AI manifests itself in MOBAs in a number of places including tactical decision making and pathfinding of minions.
Students are asked to consider a single-player version of a MOBA where a user plays against a fully-automated opponent.
Thus, AI must also be implemented for the opponent Hero's pathfinding and more complicated, strategic decision-making.


\subsection{Supporting AI Instruction with Modular Design}

\begin{figure}
\includegraphics[width=3.5in]{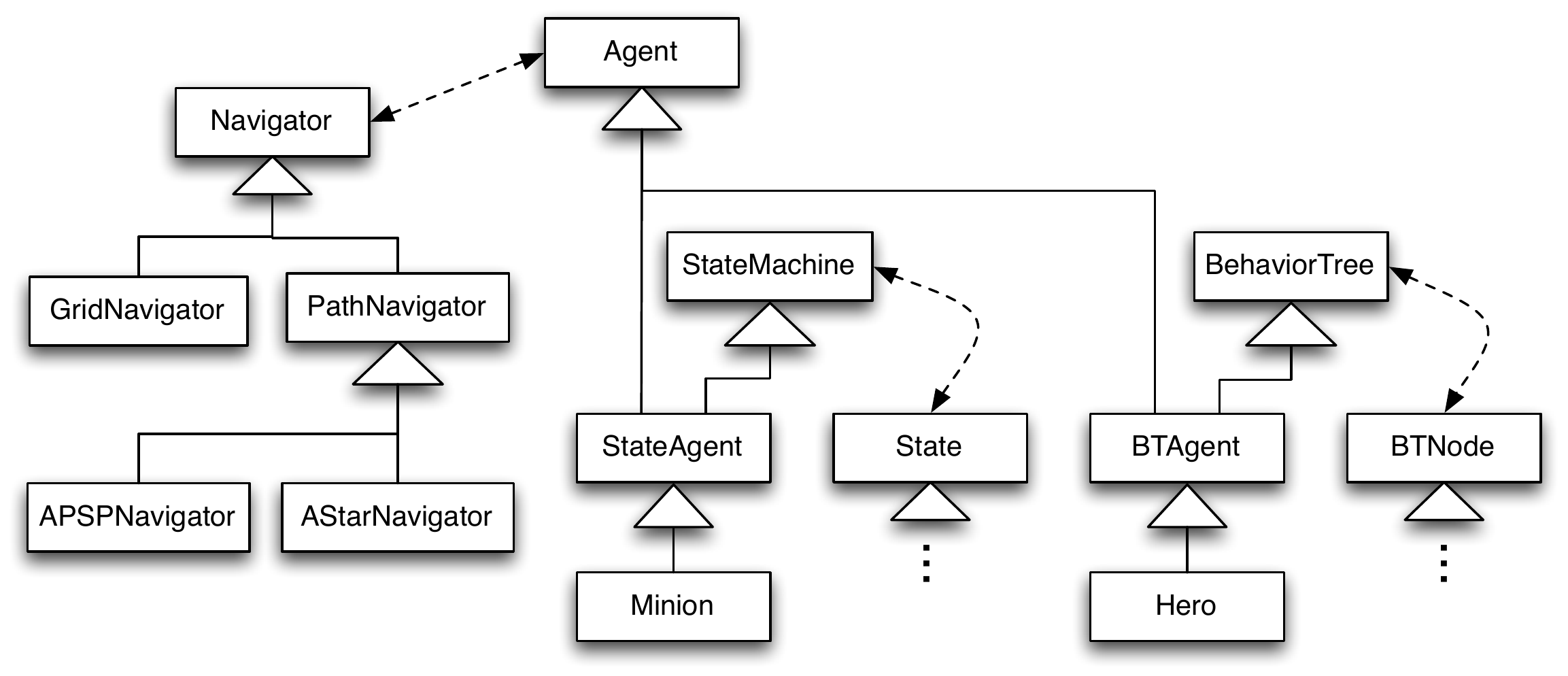}
\caption{Class diagram showing the modular design of artificial intelligence in GAIGE.}
\label{fig:uml}
\end{figure}

To facilitate Game AI instruction, GAIGE uses a modular design in which artificial intelligence is abstracted out of the agent into separate, object-oriented modules.
There are two principal classes of intelligence in computer games: (1) {\em path finding}, in which an agent must find the optimal set of points to traverse to reach a target destination without colliding with the physical terrain, and (2) {\em decision making}, in which an agent determines, at run time, which behaviors and animations to trigger (including initiating path finding). 
In GAIGE, agents are very simple and are, by themselves, incapable of performing behavior more sophisticated than turning, shooting, and walking in a straight line. 
Instead of directly modifying the code of the agent, intelligence is implemented in separate objects that attach to agent objects and make callbacks to the agent to control where it moves or what it does.
The modular separation of intelligence from the agent is shown in the class diagram in Figure \ref{fig:uml}.

All agents make use of a Navigator object that is responsible for monitoring an agent's location and make sure that the agent can move to a desired target location without colliding with obstacles. 
Given a target destination, Navigators can implement any algorithm to produce a path---a sequence of line segments guaranteed to avoid obstacles.
Since agents only know how to walk in a straight line, a Navigator instructs the agent at the appropriate times on where to move next.
Figure \ref{fig:uml} shows how Navigator can be sub-classed to implement grid-based navigation or two different path network search algorithms.

Unless the behavior of an agent is hard-coded into the agent itself, some external object must monitor the agent's local environment and make real-time decisions about behaviors to perform, animations to play, and where to move. 
Figure  \ref{fig:uml} shows two strategies for decision-making: finite-state machines and behavior trees.
The agent subclasses from a finite state machine or a behavior tree executor, respectively, and chooses which State object or Behavior (BTNode) object is currently ``in control''.
The State or BTNode in control is responsible for calling back to the agent with functions (e.g., move, shoot, etc.) or determining that another State or BTNode should be in control.
In GAIGE, States and BTNodes have special code that executes when the object first takes control of the agent, when the object cedes control of the agent, and at every frame.

\subsection{Using GAIGE: Assignments}

In this section, we describe the sequence of assignments that explore the different ways in which AI is essential for games and build a complete, working MOBA game.

\subsubsection{Assignment 1: Grid Navigation}
Some computer games, such as real time strategy games, use grid-based location of agents. 
In this assignment, students must set up the data structures for a navigation grid---a table of Booleans indicating traversable space in the map. This table is used by greedy navigator provided to the students. 
The point of the assignment is to familiarize the student with the inner working of the game engine and how Navigator objects control agents.

\subsubsection{Assignment 2: Navigation Meshes}
We turn away from grid-based navigation to the much more commonly used {\em pathnode networks}, in which invisible waypoints are positioned at key points in the virtual world.
A sparse set of waypoints and arcs indicate navigable areas in the map, although an agent can deviate from the pathnode network when engaged in combat, to pick up items, or to cut a corner.
In many games, pathnode networks are hard-coded into virtual worlds by designers.
However, for procedurally generated worlds, or worlds created by end-user content designers, the pathnode network must be generated automatically.

The optimal placement of pathnodes in a virtual environment is non-trivial.
One technique is to first generate a {\em navigation mesh}, a set of convex polygonal hulls that overlay navigable space.
An agent can move in a straight line between any two points within a convex hull without risk of collision with static terrain obstacles.
Waypoints can be placed along common edges between convex hulls, guaranteeing a navigable pathnode network.
In this assignment, students write the code to create a navigation mesh and the automatic construction of a pathnode network.

The next two assignments build off this one, implementing graph search on networks that were first generated by the student's code. 
This is an unique opportunity to practice implementing graph search on networks that are not pristine, toy graphs provided by the instructor. We feel this to be a more realistic scenario.
Should students' solutions to this assignment be flawed, they are provided with an instructor solution so they can move forward. 

\subsubsection{Assignment 3: All-Pairs Shortest-Path Navigation}
While A* is the most commonly used graph search algorithm, all-pairs shortest-path (APSP) algorithms such as the Floyd-Warshall algorithm can be used in worlds with static obstacles because the shortest path between any two waypoints will never change.
In this assignment, students pre-process a pathnode network (generated from a navigation mesh) with an APSP algorithm, producing a successor-node table.
Students then implement linear time routines to reconstruct the shortest path between any two waypoints.
An agent is provided that uses the students' navigation code to collect crystals placed around the world.

\subsubsection{Assignment 4: A* Navigation}
In this assignment, students are provided with a game world with dynamic obstacles, in the form of walls that randomly appear and disappear throughout the environment (see upper right of Figure \ref{fig:gaige}).
As with the previous assignment, an agent must collect crystals but must now account for the fact that any path may be blocked at any time. 
Students must implement the A* algorithm with replanning.

As an optional component to the assignment for extra credit, students can implement {\em smoothing} routines. 
Smoothing is a process of optimizing an agent's path through a space by skipping waypoints and/or cutting corners to minimize the overall distance the agent has to move to arrive at a target destination.
Literally following the arcs of a pathnode network results in robotic-looking behavior, and smoothing can make agent movement look more fluid and natural.

\subsubsection{Assignment 5: Finite State Machines}
In previous assignments, the agent is pre-programmed to collect crystals and the student must ensure it can get from crystal to crystal without collision.
In this assignment, students must determine what behaviors, including moving and shooting, agents must perform and when.
Students are provided with a MOBA game with only minion agents.
Bases from both teams spawn a minion every few seconds.
Students must implement a finite state machine that controls each minion to attack enemy towers and bases.
The AI implementation must be able to destroy an enemy base with fewer than a set number of minions.
Minion agents do not need to strictly reproduce typical MOBA minion behavior, and students are encouraged to get creative to see how fast they can destroy the enemy base and with how few minions.

While finite state machines are often learned early in a computer science curriculum, the finite state machines in games are closely integrated with game engines and behaviors can happen at state transition time as well as at every tick between  transitions.
This assignment is fundamentally about design---the translation of a specification for minion behavior over time into a finite state machine that operates in a real-time dynamic environment.

\subsubsection{Assignment 6: Behavior Trees}
\begin{figure}
\centering
\includegraphics[scale=0.5]{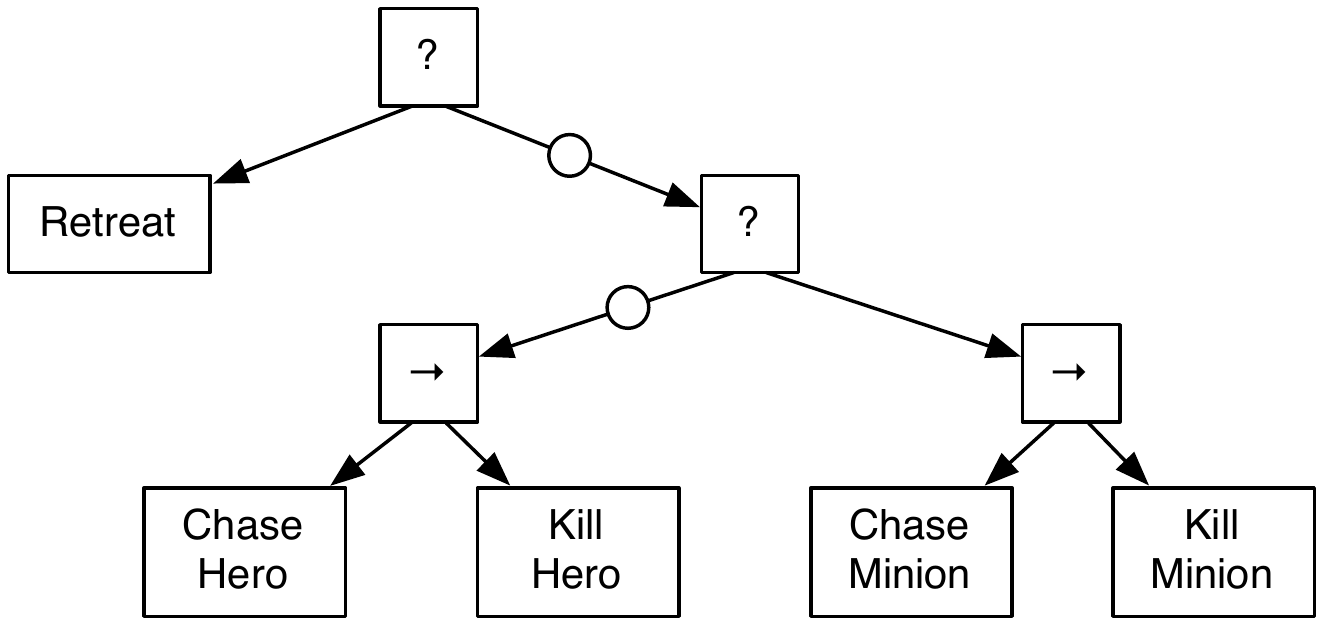}
\caption{An example of a behavior tree. 
Question-marks are {\em choice} nodes that select the first child that is applicable. 
Arrows are {\em sequence} nodes that execute all children in the order given.
}
\label{fig:behavior-tree}
\end{figure}

Whereas assignment 5 focuses on minions, assignment 6 focuses on hero agents.
Hero agents have different abilities, including two types of weapons and the ability to jump out of the way of incoming fire. 
They become stronger as they destroy opponents, but lose all enhancements when they die and respawn. 
Since there is only a single hero per team, the AI that controls a hero agent msut be more strategic. 
Students are asked to implement a behavior tree that controls a hero agent.
A behavior tree is a version of real-time hierarchical planning, capable of generating and executing temporally extended sequences of actions. 
Figure \ref{fig:behavior-tree} shows a simple behavior tree.

There are two parts to the assignment.
First, the students must complete an implementation of the behavior tree algorithm, which is tested on a suite of test inputs that does not use the game engine.
Second, the students must design a set of behaviors for use controlling a MOBA hero character.
While the behavior tree logic is simple in theory, the solution is complicated by the demands of a real-time game engine since behaviors can be temporally extended across multiple game ticks.

Students are asked to focus on hero-versus-hero combat. 
This is not a very common aspect of MOBA games.
However, by focusing on hero-versus-hero combat, the solution is not dissimilar to the AI that might be used for a first-person shooter. 
In this way, students receive practical experience with multiple game genres while still building toward a single, complete MOBA game.

\subsubsection{Assignment 7: MOBA Competition}
The final assignment is a full MOBA competition (see Figure \ref{fig:gaige}), where each student must provide the AI for minions and heroes.
Since minion AI was initially designed without consideration of the presence of heroes, and hero AI was initially designed to focus on hero-versus-hero combat, students must redesign both minion and hero AI controllers. 
This assignment is about design and the translation of design into algorithm.
Because it does not introduce any new algorithms, this assignment is optional, with extra credit awarded based on competition performance.





\subsection{Using GAIGE: Experiences}

GAIGE has been used twice, once by the author of this paper, and once by another instructor at Georgia Tech.
The first time GAIGE was used, a number of bugs were discovered in the underlying game engine, which were quickly patched.
Despite the glitches, student anonymous opinions were unanimously favorable.
Students appreciated the unified framework and the sense of building towards a larger goal. 
The course was fast-paced, with assignments every week or two.
However, students did not generally believe the course to be too difficult.
It should be noted that all students were required to have had taken an introductory AI course, which had already covered A*, thus the challenge of the A* assignment came from working with the generated path network, working with the run time environment, replanning, and smoothing.

\begin{figure}
\centering
\includegraphics[width=3in]{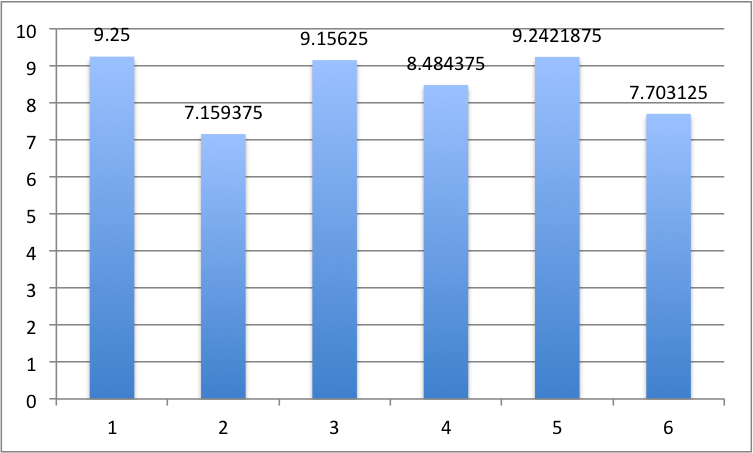}
\caption{Mean grades (out of 10 points) for each of the first six assignments.}
\label{fig:means}
\end{figure}

Assignment 2 was the hardest assignment, with a distinctly lower mean grade than the other assignments (Figure \ref{fig:means}). 
Although navigation mesh generation is relatively straight-forward, there are many edge cases that have to be handled. 
The assignment is also longer than others, having two distinct phases: navigation mesh generation and path node placement.
We have since created an Assignment 1.5, in which one must automatically create a navigation mesh from a set of pre-placed waypoints. 
The intention of this assignment is to familiarize students with path networks and how to determine which waypoints should be connected.

As seen in Figure \ref{fig:means}, students also found assignment 6 to be more challenging than other assignments.
Students' agents must beat three, instructor-provided baseline hero agents in hero-versus-hero game play.
Although the baseline agents were designed to be sub-optimal, one of the baselines turned out to be relatively difficult to beat; the game is simple enough that even a relatively simple strategy can be close to optimal.
Future work requires revisions to the assignment 6 baselines and autograder. 

The second time GAIGE was used a server was set up in which students could submit their solutions to the autograder for immediate feedback.
The introduction of assignment 1.5 improved the mean grade of assignment 2 to 7.5/10; other assignments were in line with earlier scores. 

Many students chose to rely on an instructor-provided solution to assignment 2 for subsequent assignments. 
This did not impair the performance of students since the modular nature of GAIGE meant that subsequent AI algorithms did not need to know the particulars of path network generation.
In some cases, student solutions were superior to the instructor solution. 
Some students also used an instructor-provided A* implementation for use on assignments 5, 6, and 7.
Due to the modular nature of GAIGE, AI behavior-control (finite state machines and behavior trees) do not require knowledge about path finding to work together.

GAIGE makes it easy to work with to create new assignments.
After students showed a surprising amount on interest in a lecture on binary space partitions, we were able to create a new assignment in less than a week.
A binary space partition is a binary search tree in which nodes represent rectangular portions of the game world in which obstacles appear.
The binary search tree can be used to filter the geometry of a game world when performing line-of-sight calculations, increasing computational efficiency of the game engine.
This assignment was ultimately not released to the students, but may be used in future iterations of the class.

We believe that it will be possible and relatively easy to create GAIGE structured assignments for procedural content generation, data mining and analytics, player modeling, and dynamic difficulty adjustment.
In the Spring of 2015, graduate students enrolled in our Game AI performed an additional assignment: to develop an algorithm that procedurally generated GAIGE game maps based on player metrics.
Graduate students were allowed to modify any aspect of the game engine necessary to collect statistics about human players' in-game behaviors and develop an algorithm that produced demonstrably different map configurations based on the data collected.


\section{Conclusions}

Game AI is distinct from conventional artificial intelligence, yet both draw from the same algorithmic roots.
Game AI has the additional appeal of using computer games to motivate and inspire student interest. 
Thus Game AI can act as an alternative to a standard course on artificial intelligence, or a follow-on course.
Game AI emphasizes the relationship between design and algorithm. 
The practice of artificial intelligence always involves an element of design, in the form of choice about representations, data structures, and algorithms.
This choice is often lost in artificial intelligence courses where problems are well-defined and packaged up for students to solve.
In a Game AI course, there are ample opportunities to practice making choices about representation and algorithm to solve a real problem.

GAIGE is a Python game engine designed to support a progression of assignments that implement the AI components of a Multi-User Battle Arena game.
Each assignment explores one way in which AI is used in modern computer game design and development. 
The progression breaks the task of creating a complete game into well-defined chunks and creates a sense of accomplishment.

GAIGE shows how artificial intelligence and game design can be practiced in a synchronized fashion without sacrificing well-defined metrics for success that make autograding possible.
The presence of autograders means that, with GAIGE, Game AI courses can now be scaled up to large class sizes, which is important as computer science enrollments grow and online instruction becomes more prevalent.


\bibliography{gaige.bib}
\bibliographystyle{aaai}

\end{document}